\documentclass[a4paper,aps,floats,intlimits,pre,notitlepage,intlimits,superscriptaddress,9pt,floatfix,twopages,twocolumn,8pt]{revtex4-1}
\usepackage{amsmath,dsfont,amsfonts}
\usepackage{mathtools,multirow,xfrac}
\usepackage[caption=false]{subfig}
\usepackage[english]{babel}
\usepackage{flushend}
\usepackage{hyperref}
\hypersetup{pdfauthor={Caracciolo Fabbricatore Gherardi Marino Parisi Sicuro},pdftitle={Random dimer model},%
            colorlinks, linktocpage=true, pdfstartpage=3, pdfstartview=FitV,%
    breaklinks=true, pdfpagemode=UseNone, pageanchor=true, pdfpagemode=UseOutlines,%
    plainpages=false, bookmarksnumbered, bookmarksopen=true, bookmarksopenlevel=1,%
    hypertexnames=true, pdfhighlight=/O,%
    urlcolor=orange, linkcolor=blue, citecolor=red, 
        }

\usepackage{pgfplots}

\usepackage{libertine}
\usepackage[libertine,smallerops]{newtxmath}
\usepackage[T1]{fontenc}

\usepackage[cal=esstix]{mathalfa}

\DeclareMathOperator{\Var}{Var}
\DeclareMathOperator{\Prob}{{\mathbb P}}

\newcommand{\mG}{{\mathcal G}}
\newcommand{\mV}{{\mathcal V}}
\newcommand{\mE}{{\mathcal E}}
\newcommand{\mD}{{\mathcal D}}

\newcommand{\dd}{\mathrm{d}}
\newcommand{\e}{\mathrm{e}}

\newcommand{\media}[1]{{\left\langle#1\right\rangle}}
\newcommand{\mediaE}[2][]{{\left\langle#2\right\rangle}_{#1}}

\usepackage{tikz}
\usetikzlibrary{patterns,decorations,arrows,shapes.geometric,arrows,mindmap,backgrounds,positioning,automata,positioning,fit,backgrounds,positioning,arrows,matrix,calc}

\tikzset{
  common/.style={draw,name=#1,node contents={},inner sep=0,minimum size=2},
  disc/.style={circle,common=#1},
  square/.style={rectangle,common={#1}},
}

\usepackage{todonotes}

\allowdisplaybreaks
\begin{document}
\title{Criticality and conformality in the random dimer model}
\author{S.~Caracciolo}
\address{Dipartimento di Fisica dell'Universit\`a di Milano, and INFN, sez.~di Milano,Via Celoria 16, 20100 Milano, Italy}
\author{R.~Fabbricatore}
\address{Dipartimento di Fisica dell'Universit\`a di Milano, and INFN, sez.~di Milano,Via Celoria 16, 20100 Milano, Italy}
\author{M.~Gherardi}
\address{Dipartimento di Fisica dell'Universit\`a di Milano, and INFN, sez.~di Milano,Via Celoria 16, 20100 Milano, Italy}
\author{R.~Marino}\affiliation{Laboratoire de Th\'eorie des Communications, EPFL, 1015, Lausanne, Switzerland}
\author{G.~Parisi}\affiliation{Dipartimento di Fisica, INFN -- Sezione di Roma1, CNR-IPCF UOS Roma Kerberos, Sapienza Universit\`a di Roma, P.le A. Moro 2, I-00185, Rome, Italy}
\author{G.~Sicuro}\email{gabriele.sicuro@for.unipi.it}
\affiliation{Department of Mathematics, King's College London, London WC2R 2LS, United Kingdom}
\affiliation{IdePHICS Laboratory, EPFL, 1015, Lausanne, Switzerland}
\begin{abstract}
In critical systems, the effect of a localized perturbation affects points that are arbitrarily far from the perturbation location. In this paper, we study the effect of localized perturbations on the solution of the random dimer problem in $2D$. By means of an accurate numerical analysis, we show that a local perturbation of the optimal covering induces an excitation whose size is extensive with finite probability. We compute the fractal dimension of the excitations and scaling exponents. In particular, excitations in random dimer problems on non-bipartite lattices have the same statistical properties of domain walls in the $2D$ spin glass. Excitations produced in bipartite lattices, instead, are compatible with a loop-erased self-avoiding random walk process. In both cases, we find evidence of conformal invariance of the excitations that is compatible with $\mathrm{SLE}_\kappa$ with parameter $\kappa$ depending on the bipartiteness of the underlying lattice only.
\end{abstract}
\maketitle
\section{Introduction}
Let us suppose that we are given a graph $\mG$. A \textit{dimer} on $\mG$ is an edge of $\mG$ with its endpoints, which are said to be ``covered'' by the dimer. A dimer covering of $\mG$ is a subset of its edge set such that each vertex is covered by one dimer only. In other words, a dimer covering is a perfect matching on the graph $\mG$ \cite{lovasz2009matching}. Despite this apparently abstract definition, dimer coverings appear in different models and theories in statistical physics. Deposition of diatomics molecules, defects in crystals, or simple magnetic systems, are all examples of problems that can be studied as dimer models \cite{Kasteleyn1967}.

For this reason, the properties of dimer coverings of a given graph have received considerable attention. In the 1960s, Kasteleyn \cite{Kasteleyn1961,Kasteleyn1967} and, independently, Temperley and Fisher \cite{Temperley1961} computed the asymptotic expressions of the number of dimer coverings on infinite lattices in $1D$ and $2D$. Their results are at the basis of the Fisher--Kasteleyn--Temperley algorithm for counting perfect matchings in planar graphs. Remarkably, they also made clear the correspondence between the solution of a dimer covering problem and the solution of the $2D$ Ising model on a lattice \cite{Kasteleyn1963, Kasteleyn1967}. 

In the present paper, we study a disordered version of the dimer covering problem, namely the random dimer model (RDM). In the RDM, covering an edge by a dimer costs a (fixed) edge-dependent random price, so that a total cost of the covering is the sum of the costs of the covered edges. In consequence, there is an \textit{optimal} way of performing a covering. The nontrivial ``optimal configuration'' can be thought of as a ``ground state'' of a disordered system. In this paper, we will focus in particular on dimer coverings of $2D$ lattices and we will show that there is a correspondence between the RDM and glassy systems in $2D$. Unlike the $2D$ Ising model, that is critical at finite temperature and has trivial ground state, the RDM has a nontrivial ground state and a critical behavior exactly at ``zero temperature'', in analogy with the physics of $2D$ spin glasses. Moreover, we will give evidences of conformal invariance of the optimal solution, a not obvious property that we will relate to observed conformality of $2D$ spin glasses at zero temperature.

More specifically, we assume that a graph $\mG(\mV, \mE)$ is given, with vertex set $\mV$, $|\mV|=2N$, and edge set $\mE\subset\mV\times\mV$. A weight $w_{e}$ is associated with each edge $e\in\mE$ of the lattice. We assume that the weights $w_e$ are independently and identically distributed random variables, having an absolutely continuous probability density $\varrho(w)$. We also assume that the graph admits more than one dimer covering. We can assign a cost $E[\mD]$ to each covering $\mD$ as
\begin{equation}
E[\mD]\coloneqq \sum_{e\in\mD}w_{e},
\end{equation}
and a corresponding Gibbs weight $\e^{-\beta E[\mD]}$, depending on the fictitious inverse temperature $\beta$. The associated partition function is given by
\begin{equation}
Z(\beta)\coloneqq\sum_{\mD}\e^{-\beta E[\mD]}.
\end{equation}
In the $\beta\to 0$ limit, all dimer coverings have the same weight and $Z(0)$ is simply the number of coverings on $\mG$. The computation of $Z(0)$ when $\mG$ is a planar graph has been the object of intense study since the seminal results of Kasteleyn, Fisher, and Temperley, recently extended to oriented surfaces \cite{Cimasoni2007,*Cimasoni2008}. We say that $\mD^*$ is an optimal covering if
\begin{equation}
E[\mD^*]=\min_{\mD}E[\mD]=-\lim_{\beta\to+\infty}\beta^{-1}\ln Z(\beta).
\end{equation}
The optimal configuration is almost surely unique and corresponds to the ``ground state'' of the model. The introduced randomness can be thought as due to noise, or impurities. In a dimer deposition picture, for example, randomness in edge weights might refer to a space-dependent binding energy of the diatomic molecules on the substrate. The RDM has been also used as a model for disordered quantum magnets \cite{Kimchi2018}.

If $\mG$ is a complete graph or a random graph, the RDM recovers the ``random-link models'' studied by statistical physicists since the 1980s \cite{Orland1985,Mezard1985,Parisi2002,Caracciolo2017,Perrupato2020}. 

As we will show below, the RDM is also related to random Euclidean matching problems (REMPs) \cite{Mezard1986a,Mezard1988}. In a REMP, a set of $2N$ points $\{ x_{i}\}_{i=1}^{2N}$ is given, uniformly and independently generated on a given Euclidean domain in $D$ dimensions (e.g., the unit hypercube). The goal is to find the optimal permutation of $2N$ elements $\sigma$ that minimizes the cost
\begin{equation}
E[\mD_\sigma]\coloneqq\sum_{i=1}^N d^p(x_{\sigma(2i-1)},x_{\sigma(2i)}),\quad p\in\mathds R^+,
\end{equation}
where $d( x, y)$ is the Euclidean distance between $x$ and $ y$. Each permutation defines a pairing, i.e., a set of matched points $\mD_\sigma\coloneqq\{(x_{\sigma(2i)},x_{\sigma(2i-1)})\}_{i=1}^N$. We denote $\mD^*\equiv \mD_{\sigma^*}$ with $\sigma^*=\arg\min_\sigma E[\mD_\sigma]$. In a \textit{bipartite} variation of the problem, the random Euclidean assignment problem (REAP), two sets of $N$ random points of distinct ``colors'', $\{ x_{i}\}_{i=1}^{N}$ and $\{y_{i}\}_{i=1}^{N}$, are given. The pairing has to be such that only points of different color are matched. In other words, we search for an optimal permutation of $N$ elements $\sigma$ that minimizes the cost
\begin{equation}\label{costoeu}
E[\mD_\sigma]\coloneqq\sum_{i=1}^N d^p(x_{i},y_{\sigma(i)}),\quad p\in\mathds R^+.
\end{equation}
Here each permutation defines a set of $N$ pairs $\mD_\sigma\coloneqq\{(x_i,y_{\sigma(i)})\}_{i=1}^N$ of points from different sets that are matched. Computing the average optimal cost in the REMP or in the REAP is a challenging task, due to the presence of Euclidean correlations between the pair costs \cite{Mezard1988,Caracciolo2014,Caracciolo2015,Lucibello2017,Caracciolo2017b}. The aforementioned random-link models provide the infinite-dimensional limit of REMPs and REAPs \cite{Mezard1988}. 

The typical properties of the solutions of these problems are not trivial. It has been shown, for example, that the random-link model on the complete graph is ``critical'', i.e., its Hessian spectrum is gapless \cite{Lucibello2017}. In Ref.~\cite{Boniolo2014}, evidence of long-range correlations in the solution of the $1D$ REAP was given. On the other hand, the RDM in the $\beta\to 0$ limit on planar graphs showed interesting conformality properties. For example, given two \textit{uniformly sampled} dimer coverings, their union generates a set of curves and paths. Kenyon predicted the convergence of such curves to a Schramm-L\"owner evolution ${\rm SLE}_4$ \cite{Kenyon2011}, and proved the conformality of the loops \cite{Kenyon2014} on bipartite lattices. He also showed that the limit measure of the possible dimer coverings has conformal invariance properties on the square lattice \cite{Kenyon2000}. We recall here that an $\mathrm{SLE}_\kappa$ curve $\gamma(t)$ in the upper complex plane $\mathbb H$ is given by $\gamma(t)=g_t^{-1}(\xi_t)$, where $g_t(z)$ satisfies the L\"owner equation,
\begin{equation}
\frac{\dd g_t(z)}{\dd t}=\frac{2}{g_t(z)- \xi_t},\qquad \frac{g_0(z)}{z}=\lim_{z\to+\infty}\frac{g_t(z)}{z}=1.
\label{loew}
\end{equation}
Here $\xi_t$ is the driving function of the process, and in the case of a $\mathrm{SLE}_\kappa$ it is given by a Brownian process with $\langle\xi_t\rangle=0$ and $\langle\xi^2_t\rangle=\kappa t$ \cite{Rohde2005,Cardy2005}.

Finally, it is known that there is a special correspondence between matchings and spin glasses in two dimensions \cite{Kasteleyn1963,fisher1966}. The problem of finding the ground state of the two-dimensional Edwards--Anderson (EA) model \cite{Edwards1975} can be mapped into a planar matching problem \cite{Bieche1980,Barahona1982}. 
Notably, the $2D$ EA model has a glass transition at zero temperature exhibiting scale invariance \cite{Bray1984,Shirakura1997,Hartmann2001}, and it has been suggested that conformal invariance might hold as well. In particular, EA domain walls have been found to be consistently described by a $\mathrm{SLE}_\kappa$ with $\kappa\approx 2.1$ \cite{Amoruso2006,Bernard2007}. 

These results motivated us to investigate the presence of criticality and conformality in the RDM on $2D$ lattices in the $\beta\to+\infty$ limit, i.e.~on the ground state, and search for correspondences with the REMP, the REAP and $2D$ spin glasses. We will show that similar critical properties appear in some RDMs, in the REMP, and in the EA model, suggesting the existence of a unique universality class for these models. Moreover, such properties depend on the nature of the underlying graph only.

\section{Models and methods.} 
\begin{figure}
{\centering \includegraphics[width=0.614\columnwidth]{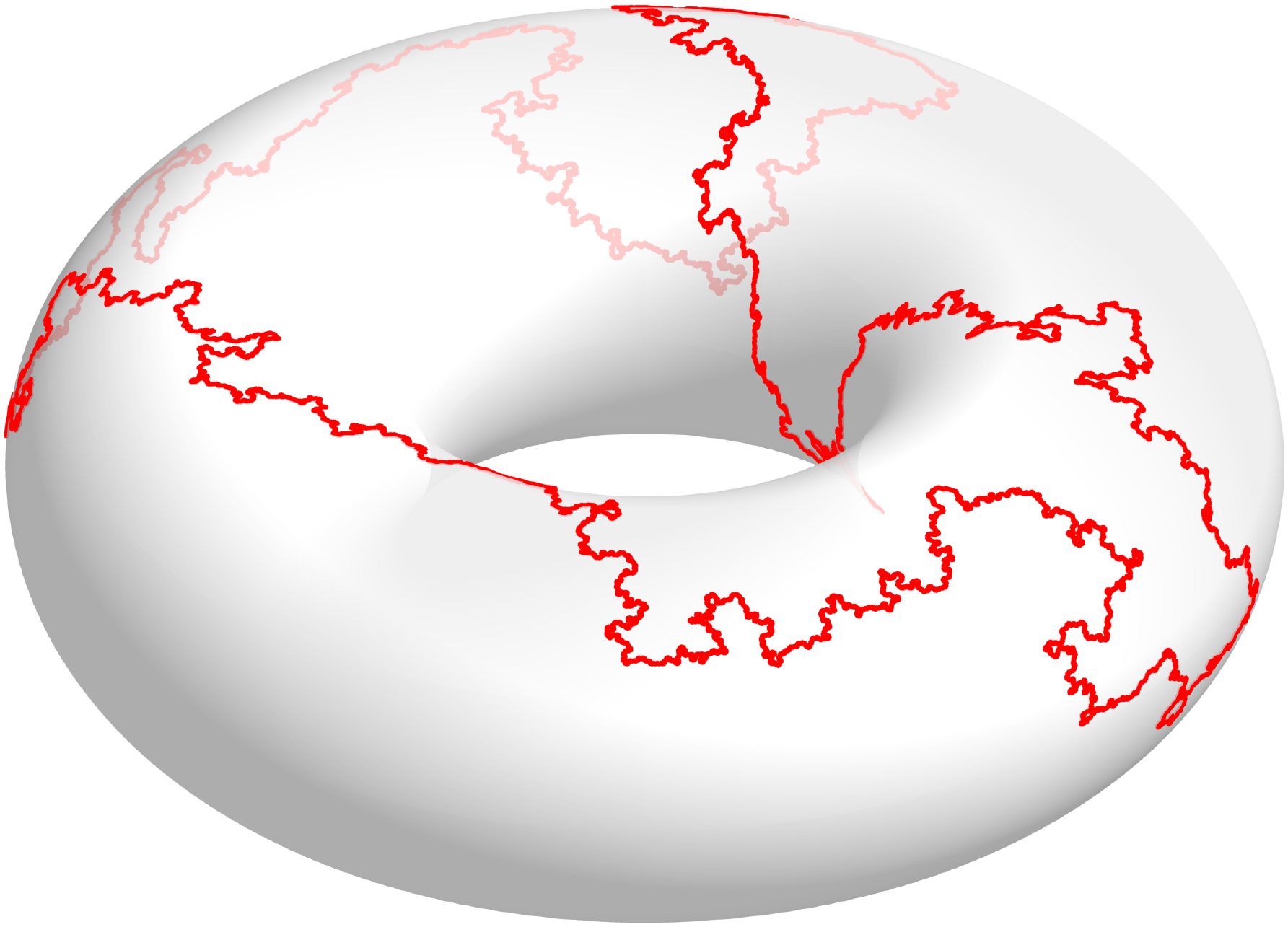}}
\caption{Example of cycle obtained comparing the minimum cost solution and an excited random solution in the random dimer covering on the honeycomb lattice with $2N=10^6$, assuming periodic boundary conditions. In this case, the length of the cycle is $S=10746$.
\label{fig:toro}}
\end{figure}

We analyze three types of lattices: the honeycomb (H) lattice, the triangular (T) lattice, and the square (Q) lattice. Each lattice is obtained considering $L$ rows of $L$ sites, displaced in such a way that the lattice edge length is fixed to $1$. The total number of sites is therefore $2N=L^2$. We impose periodic boundary conditions in both directions. For $L=4$, for example, the three lattices are
\begin{center}
\includegraphics[width=\columnwidth]{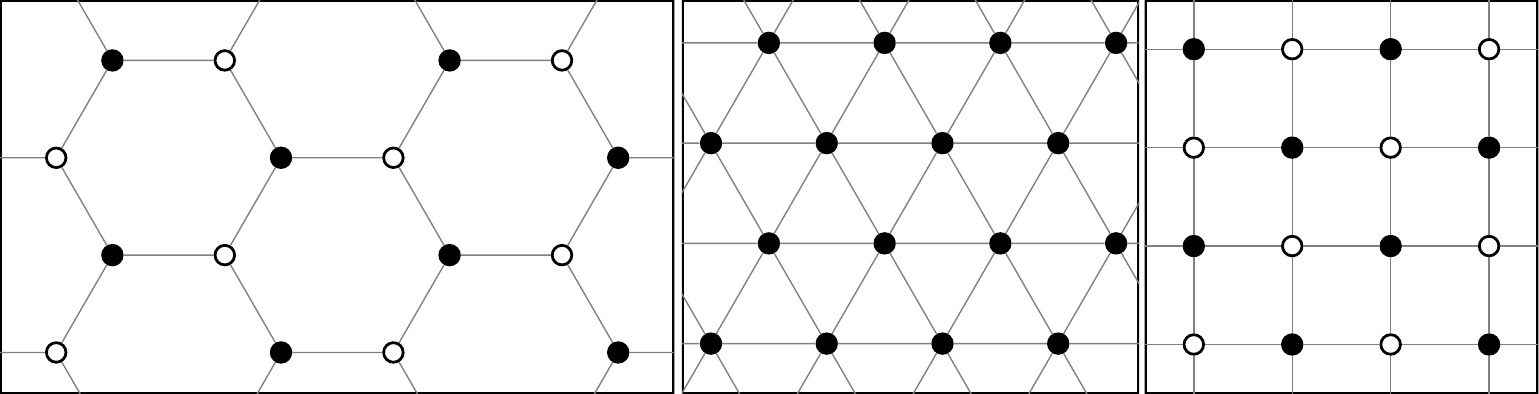}
\end{center}
We associate to each edge $e$ a random weight $w_e$, extracted from the exponential distribution $\varrho(w)=\e^{-w}$. Once all weights have been assigned, Edmond's blossom algorithm \cite{Edmonds1965,lemon} provides us the optimal weighted dimer covering $\mD^*$. Subsequently, we select a random edge $\hat e\in\mD^*$ and we cut it. Forbidding the edge $\hat e$ plays the role of a local perturbation that induces an excitation. Re-running the algorithm on the graph in which $\hat e$ is not present, we obtain a new optimal solution $\mD_{\hat e}^*$ of higher cost with respect to $\mD^*$. The difference $\Delta E_{\hat e}\coloneqq E[\mD_{\hat e}^*]-E[\mD^*]\geq 0$ scales as $\Delta E_{\hat e}=\mathcal O(1)$ for $N\gg 1$. To evaluate the extent of the perturbation effect, we consider the symmetric difference between $\mD_{\hat e}^*$ and $\mD^*$:
\begin{equation}\label{ciclo}
\mathcal S_{\hat e}=\{e\in \mathcal{ E}\colon e\in \mD^*\triangle\mD_{\hat e}^*\}.
\end{equation}
This selects a set of edges in the original graph. Pictorially, e.g., on the H model
\begin{center}
\includegraphics[width=\columnwidth]{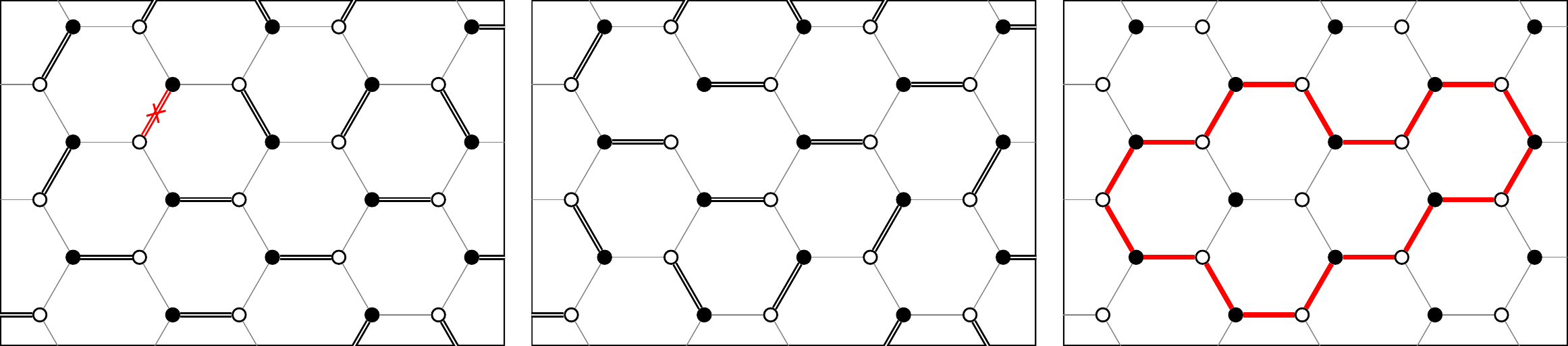}
\end{center}
Obviously, $\hat e\in \mathcal S_{\hat e}$ and it is easy to see that $\mathcal S_{\hat e}$ is a self-avoiding single cycle. 
 If the system is critical, we expect that the localized perturbation induces, with finite probability, an extensive rearrangement that affects arbitrarily far points. The cycle $\mathcal S_{\hat e}$ is a fractal object; see Fig.~\ref{fig:toro}. We will denote by $S_{\hat e}\coloneqq|\mathcal S_{\hat e}|$ the number of edges of the cycle, thus its size.

Alongside the RDM, we consider the REMP and the REAP with a total of $2N$ points on a square of size $L=\sqrt{2N}$ with a periodic boundary condition. We adopt the cost in Eq.~\eqref{costoeu} with $p=2$ \footnote{In Ref.~\cite{Mezard1985} it is shown that the $p=D$ case has an exactly solvable $D\to+\infty$ limit. We focused therefore on $p=D=2$.}. We study the excitations in these problems exactly as in the RDM, exciting the optimal solution $\mD^*$ into a new matching $\mD^*_{\hat e}$ by forbidding a pair $\hat e\in\mD^*$. In the REMP, the difference between $\sigma^*$ and $\sigma^*_{\hat e}$ can be quantified as before considering the cycle $\mathcal S_{\hat e}$ as in Eq.~\eqref{ciclo}. In the REAP the cycle $\mathcal S_{\hat e}$ is obtained starting from the edge set in Eq.~\eqref{ciclo} and then joining consecutive points of the same color: this is because in the REAP the typical distance of a pair in the optimal solution scales as $\sqrt{\ln N}$ \cite{Ajtai1984}, whereas the distance between points of the same color is $\mathcal O(1)$ as in all other considered cases.

\section{{Criticality and fractal dimension.}} 
\begin{figure}
\subfloat[]{\includegraphics[height=0.9\columnwidth]{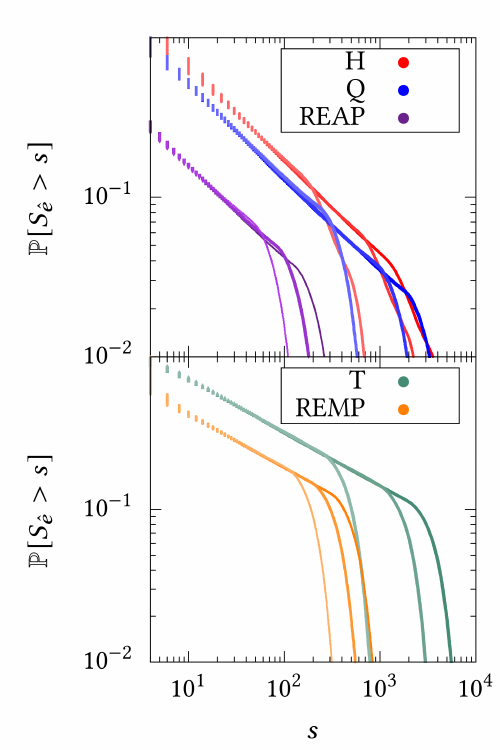}}\kern-1em
\subfloat[]{\includegraphics[height=0.855\columnwidth]{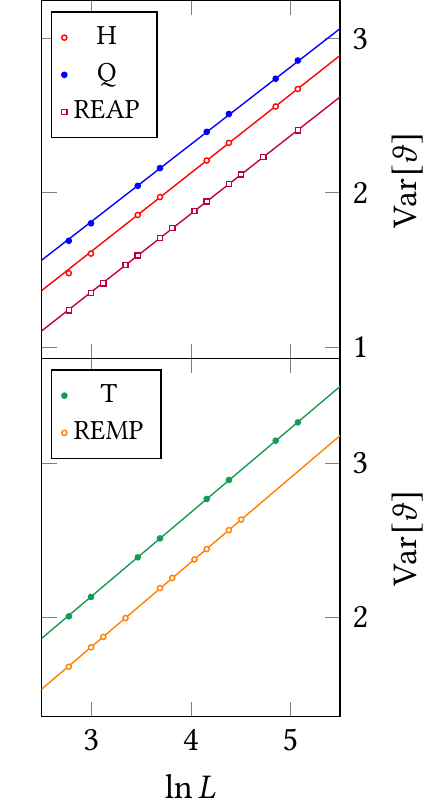}}
\caption{(a) Cumulative distribution of the size $s$ of the excitation for the different models considered in the paper. The represented sizes are $L=100,300,500$ (from left to right in each case) for each dimer model. For the Euclidean cases, instead, the represented sizes are $2N=2000,5000,10000$, from left to right. (b) Variance of the winding angle as a function of $L$ for all models. Lines are fits obtained using the function $f(L)=a+\frac{\kappa}{4}\ln{L}$.}
\label{isto}
\end{figure}

We numerically evaluated the probability $\Prob[S_{\hat e}>s]$ of having a cycle of length greater than $s$ for all models considered above. Scaling theory for critical systems states that such a probability can be written as
\begin{equation}\label{plansatz}
\Prob[S_{\hat e}>s]=s^{-\zeta}\rho\left(s^\lambda L^{-1}\right),
\end{equation}
for some scaling function $\rho$, such that $0<\lim_{z\to 0}\rho(z)<+\infty$. The scaling exponents $\zeta>0$ and $\lambda>0$ have to be determined. The scaling ansatz in Eq.~\eqref{plansatz} is numerically confirmed for all the analyzed cases, see Fig.~\ref{isto}. As the size $L$ increases, a power-law tail develops in all considered models. This implies that local perturbations induce extensive rearrangements with finite probability in the thermodynamic limit and the models are indeed on a ``critical point''. A numerical estimation of $\zeta$ by fitting the tail in Fig.~\ref{isto} also shows that the power-law exponent is the same for the H and the Q model and the REAP, whereas the T model has a different exponent very close to the REMP one; see Table~\ref{tab:esponenti}.

\begin{figure}
\includegraphics[width=\columnwidth]{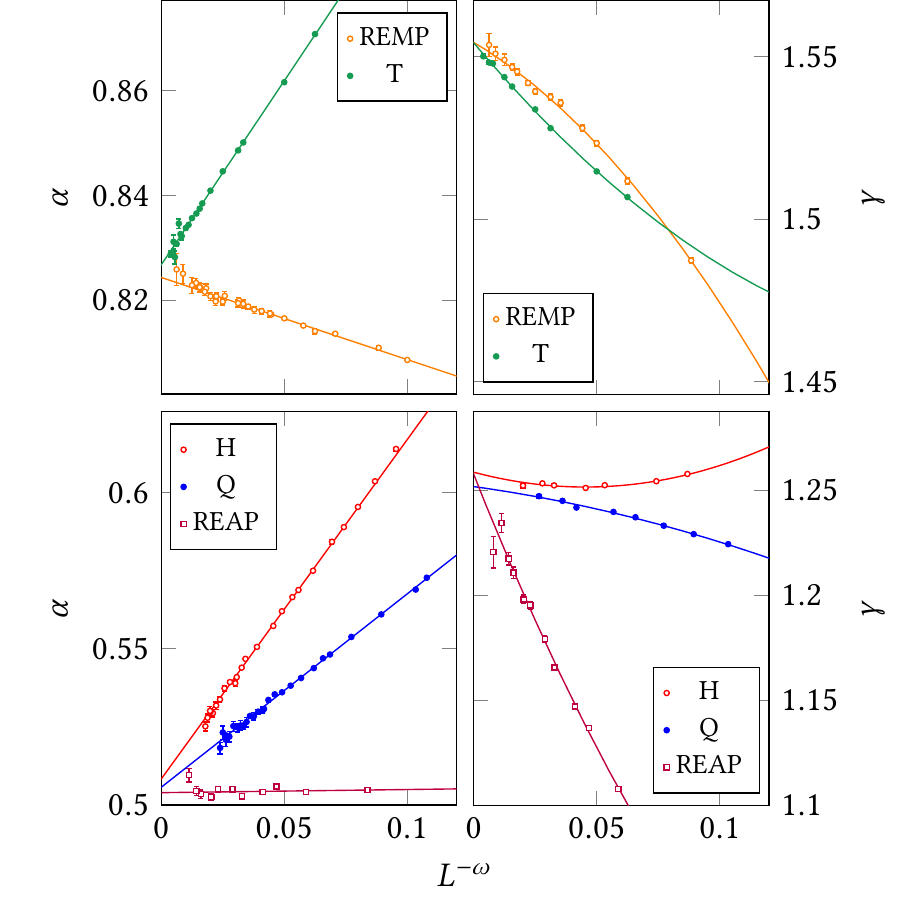}
\caption{Extrapolation of $\alpha$ (left) and $\gamma$ (right) for the different models. The value of both exponents in the bipartite models is clearly different from the one obtained for the monopartite ones.\label{fig:alphagamma}}
\end{figure}

Let us now evaluate the fractal dimension $D_{\mathrm f}$ of the cycle. Assuming now that $\media{S_{\hat e}} \sim L^\alpha$, where $\media{\bullet}$ is the average over all instances, from Eq.~\eqref{plansatz} we have
\begin{equation}\label{alpha}
    \alpha=\frac{1-\zeta}{\lambda}.
\end{equation}
The gyration radius of the cycle is defined as
\begin{equation}
    R^2_{\hat e}\coloneqq\frac{1}{2S^2_{\hat e}}\sum_{i,j}d^2( r_i, r_j),
\end{equation}
where $ r_i$ is the position of the $i$th node in the cycle and the sum runs over all pairs of vertices of the cycle. Conditioning on cycles of length $s$, it satisfies a scaling law of the form \footnote{Here a hyperscaling assumption has been made: we assume that, to get a proper $L\to+\infty$ limit, the argument of $g$ in Eq.~\eqref{r2ansatz} has the same scaling properties of the one of $\rho$ in Eq.~\eqref{plansatz}. }
\begin{equation}\label{r2ansatz}
    \mediaE[S_{\hat e}=s]{R^2_{\hat e}}=s^{2D_{\mathrm f}^{-1}}g\left(s^\lambda L^{-1}\right),
\end{equation}
where $g$ is a scaling function such that $0<\lim_{z\to 0}g(z)<+\infty$. Assuming that $\media{R^2_{\hat e}}\sim L^\gamma$, we obtain
\begin{equation}
    \gamma=\frac{2D_{\mathrm f}^{-1}-\zeta}{\lambda}.
\end{equation}
With reference to Eq.~\eqref{alpha} a relation between the three exponents $\alpha$, $\gamma$ and $\zeta$ and the fractal dimension $D_{\mathrm f}$ is easily found:
\begin{subequations}\label{DZ}
\begin{align}
D_{\mathrm f}&=2-\gamma+\alpha,\label{df_rel}\\
\zeta&=\frac{2-\gamma}{2-\gamma+\alpha}\label{zeta_rel}.
\end{align}
\end{subequations}
The fractal dimension and the power-law exponent $\zeta$ can be extracted by a careful measurement of $\alpha$ and $\gamma$. These quantities can be estimated using the method of ratios \cite{Caracciolo1995}, i.e., considering, for each $L$, a system of size $L$ and a system of size $2L$. Assuming now that $\langle S_{\hat e} \rangle\sim L^\alpha$, the value of the exponent $\alpha$ has been estimated as
\begin{equation}
\log_2\frac{\mediaE[2L]{S_{\hat e}}}{\mediaE[L]{S_{\hat e}}}=\alpha+\frac{\alpha^{(1)}}{L^{\omega}}+o\left(\frac{1}{L^{\omega}}\right),
\end{equation}
where $\mediaE[L]{\bullet}$ denotes an average at size $L$. The fit of our data, given in Fig.~\ref{fig:alphagamma}, has been performed using a fitting function $f(L)=\alpha+\alpha^{(1)}L^{-\omega}$, with $\alpha$, $\alpha^{(1)}$, and $\omega$ free parameters. Similarly for the gyration radius, assuming that $\langle R^2_{\hat e} \rangle\sim L^\gamma$, we have
\begin{equation}
\log_2\frac{\mediaE[2L]{R^2_{\hat e}}}{\mediaE[L]{R^2_{\hat e}}}=\gamma+\frac{\gamma^{(1)}}{L^{\omega}}+\frac{\gamma^{(2)}}{L^{2\omega}}+o\left(\frac{1}{L^{2\omega}}\right).
\end{equation}
In this case, the fit has been obtained using a fitting function $f(L)=\gamma+\gamma^{(1)}L^{-\omega}+\gamma^{(2)}L^{-2\omega}$, with $\gamma$, $\gamma^{(1)}$ and $\gamma^{(2)}$ free parameters, while $\omega$ was fixed to the same value estimated in the analysis for $\alpha$. The data points have been obtained averaging over $10^7$--$10^8$ different instances for each value of $L$. In the case of lattice models, we used $8\leq L\leq 512$, whereas for the REMP and the REAP we considered $64\leq L^2\leq 51200$. Our fits are given in Fig.~\ref{fig:alphagamma}.

\begin{table}\centering
\resizebox{\linewidth}{!}{
\begin{ruledtabular}
\begin{tabular}{@{}clllll@{}}
&H &Q&REAP&T&REMP\\
\colrule
$\ \lim_{N} \media{N^{-1}E[\mD^*]}$&$0.703(1)$&$0.355(1)$&$\infty$& $0.529(1)$&$0.625(1)$\\
$\media{\Delta E_{\hat e}}$&$1.115(1)$&$0.655(1)$&$0.637(1)$&$0.380(1)$&$0.585(1)$\\
$\alpha$&$0.508(2)$& $ 0.506(1)$&$0.504(1)$&$0.827(1)$&$0.824(1)$\\
$\gamma$&$1.259(1)$& $1.252(2)$&$1.257(15)$&$1.554(1)$&$1.554(1)$\\
$\omega$&$0.74(2)$& $0.65(2)$&$1.0(2)$&$1.00(2)$&$0.94(7)$\\
$\zeta$&$0.593(1)$& $0.597(1)$&$0.596(6)$&$0.350(1)$&$0.351(1)$\\
$\zeta$ (from fit)&$0.593(1)$&$0.595(1)$&$0.588(1)$&$0.354(1)$&$0.350(1)$\\
$D_{\mathrm f}$&$1.250(2)$&$1.253(2)$&$1.247(15)$&$1.273(1)$&$1.270(1)$\\
$\kappa$&$2.034(6)$&$2.003(7)$&$2.023(4)$&$2.181(4)$&$2.195(4)$\\ 
$1+\frac{\kappa}{8}$&$1.254(1)$&$1.250(1)$&$1.253(1)$&$1.273(1)$&$1.274(1)$
\end{tabular}
\end{ruledtabular}
}
\caption{Asymptotic average optimal cost and of the scaling exponents for random dimer covering problems on the torus. Note that the REAP has an anomalous scaling in the cost, namely, $\langle E[\mathcal D^*]\rangle\sim \frac{N}{\pi}\ln N$ for $N\gg 1$.} 
 \label{tab:esponenti}
\end{table}
All our results are collected in Table~\ref{tab:esponenti}. We observe that the values of the exponents naturally splits in two groups: the first one including the Q model, the H model, and the REAP, with cycles having $D_{\mathrm f}=1.252(2)$, and the second one including the T model and the REMP, with cycles having $D_{\mathrm f}=1.273(2)$. The fact that the REAP and the H and Q models share the same exponents, and similarly the T model shares its exponents with the REMP, suggests that the only relevant feature that determines the scaling is the nature of the underlying graph, i.e., the fact of being bipartite or not. In all cases $\zeta>0$, i.e., the local perturbation induces a long-range rearrangement with finite probability in the large $N$ limit. Moreover, the estimated value for $\zeta$ is in agreement with the result of the direct fit, confirming the consistency of our scaling ansatz.

\section{{Conformality.}}\label{sec:conf} 
\begin{figure*}
    \includegraphics[width=0.9\textwidth]{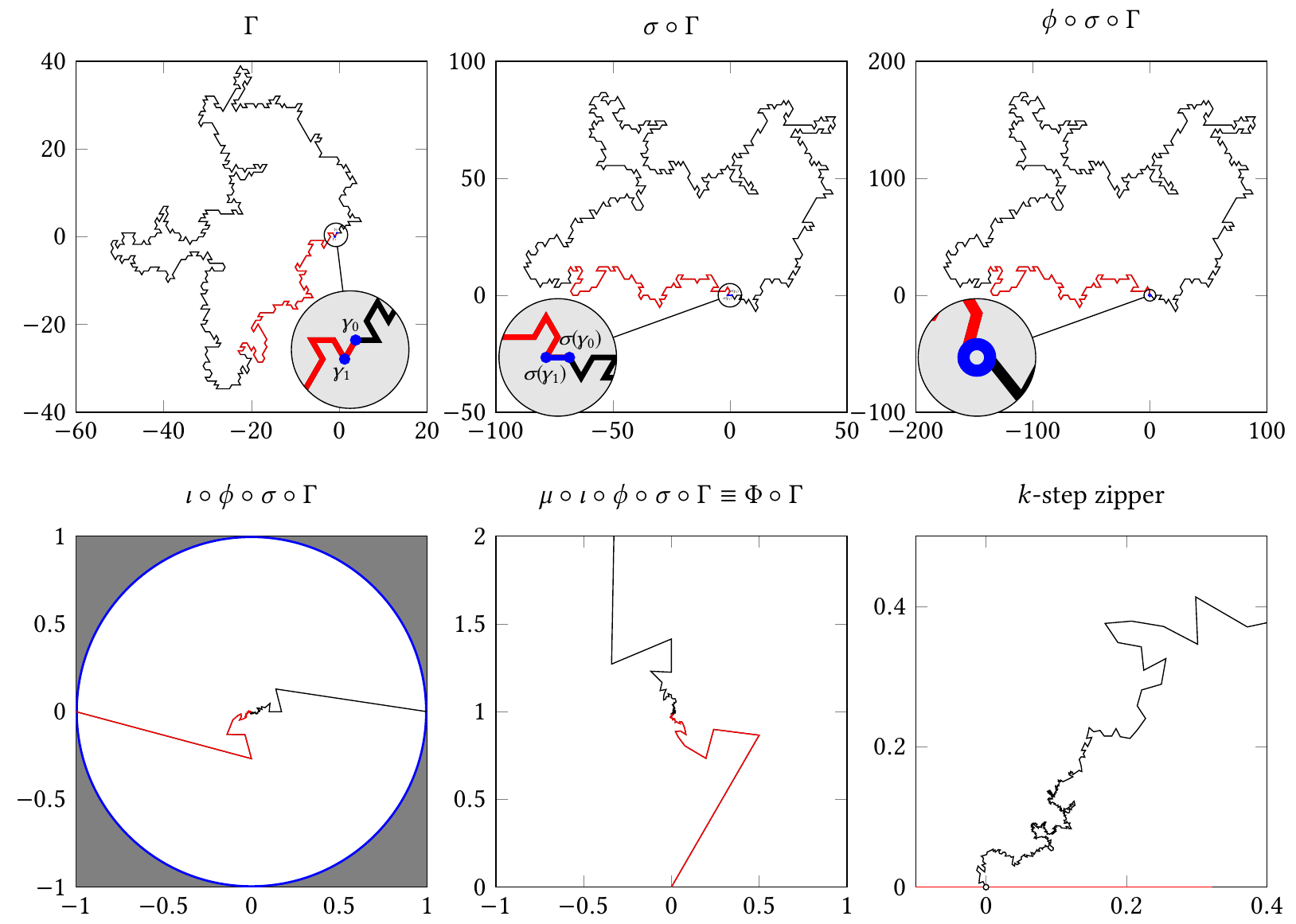}
    \caption{\label{fig:zip}A contractible loop with $s=522$ in the T model with $L=400$ is split into two parts (red, of $k=90$ steps, and black, of $s-k$ steps). By a sequence of conformal transformations, the black part is mapped into a curve in the upper half plane stemming from the origin. }
\end{figure*}

\begin{figure*}
    {\centering
    \subfloat[]{\includegraphics[height=0.6\textwidth]{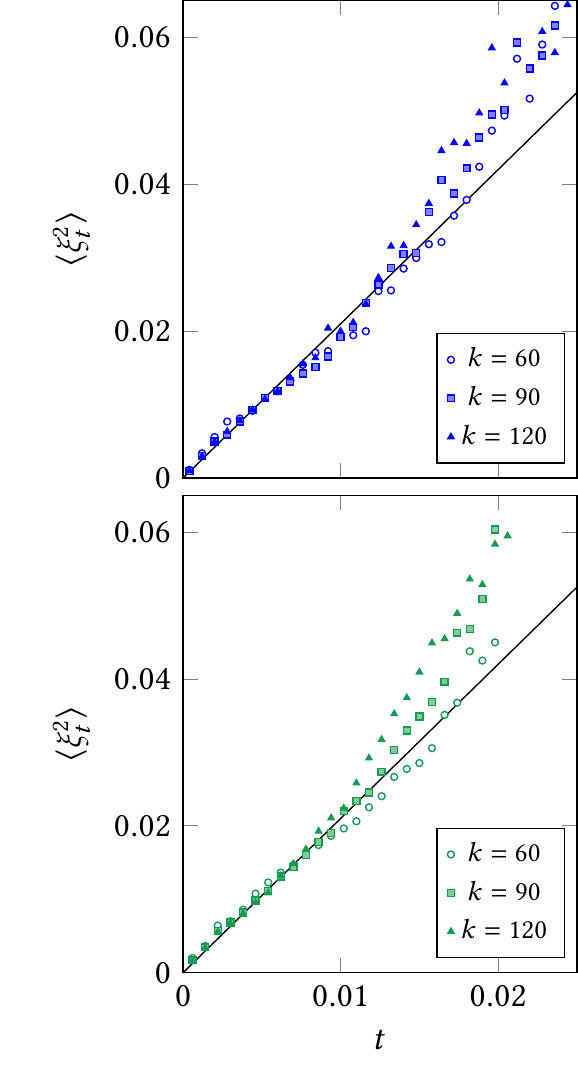}}
    \subfloat[]{\includegraphics[height=0.6\textwidth]{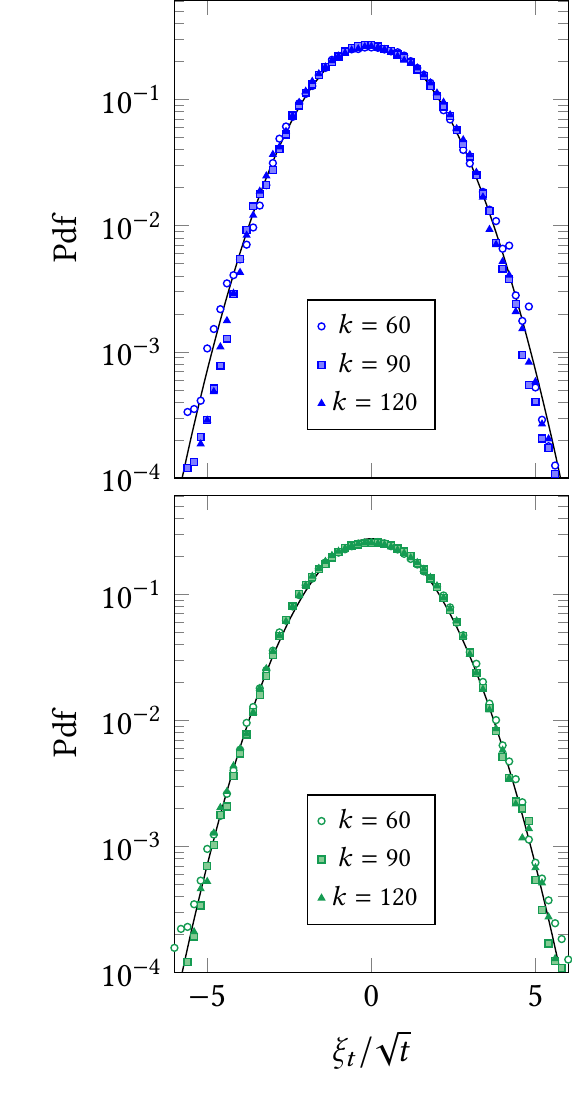}}
    \subfloat[]{\includegraphics[height=0.6\textwidth]{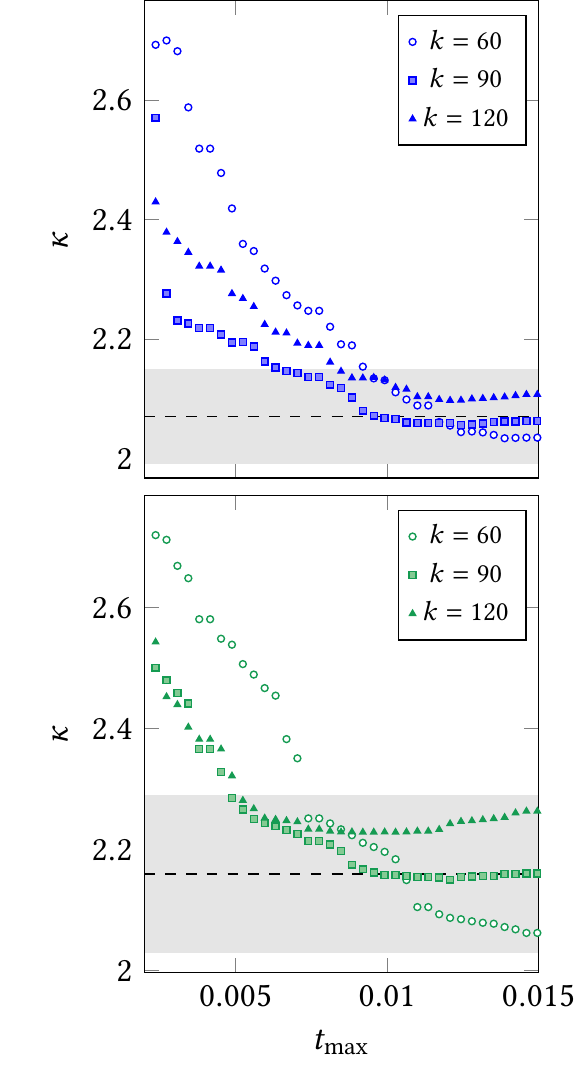}}
    }
    \caption{Results obtained from the study of the driving function on the Q lattice (top) and on the T lattice (bottom) with $L=400$ for three different values of $k$. (a) Mean square displacement as a function of time, binned in intervals $\delta t=0.0008$. (b) Probability density function of the driving function. The numerical results have been obtained averaging over a wide range of time $t$ and are compared with a normal distribution (continuous line) having zero mean and variance $\kappa=2.1$. (c) Fitted values of $\kappa$ as functions of the upper cutoff $t_\mathrm{max}$. Shaded bands indicate our final estimates (dashed line) and their variability.}
    \label{fig:forzante}
\end{figure*}

The presence of criticality suggests the inspection of a stronger invariance, namely conformal invariance. The possible presence of conformal invariance in the REAP has been suggested in Ref.~\cite{Zarinelli2008}. We restrict ourselves to the lattice models and we consider cycles $\Gamma=\left(\gamma_0,\gamma_1,\ldots,\gamma_s\equiv \gamma_0\right)$ of length $s$ on the torus. By means of the notation above, we denote the piecewise linear curve $\Gamma$ passing through $\gamma_0,\gamma_1,\ldots,\gamma_{s}\equiv\gamma_0$, $\gamma_i$ being a site on the lattice for $i=0,\dots, s$. We consider contractible cycles only. We say that a cycle is contractible if it does not wind around the torus. Fixing $k<s$, in our approach, we map a portion of the cyles into the standard chordal geometry as follows \cite{Gherardi2013}. 
\begin{enumerate}
    \item By means of a translation, rotation, and scale transformation that we denote $\sigma$, we map $\Gamma$ into a curve in $\mathbb C$ such that $\sigma(\gamma_0)=1$ and $\sigma(\gamma_1)=-1$.
    \item We map the complement in $\mathbb C$ of the real segment $[-1,1]$ to the complement in $\mathbb C$ of the unit disk, by $\phi(z)=z+\sqrt{z^2-1}$.
    \item We map the exterior to the interior of the unit disk, by $\iota(z)=\sfrac{1}{z}$.
    \item We map the interior of the disk to the upper half-plane $\mathbb H$, by a M\"obius transformation
    \begin{equation}
    \mu(z)=i\frac{1+z}{1-z}.
    \end{equation}
    \item Given now the curve $
        \tilde\Gamma=\Phi\circ\Gamma$, where $\Phi\coloneqq \mu\circ\iota\circ\phi\circ\sigma$, we perform $k$ steps of the standard zipper algorithm in $\mathbb H$. This uniformizes the first $k-1$ steps of $\tilde\Gamma$, mapping each of the $\gamma_i$ for $i=0,\dots,k-1$ into the real axis. As a result, the curve $\hat\Gamma_k=\left(\gamma_k,\gamma_{k+1},\ldots,\gamma_s\right)\subset\Gamma$ (a chordal curve from the point $\gamma_k$ to the point $\gamma_0$ in the original domain), is mapped in a curve in the upper half plane.
    \item Finally, the zipper algorithm \cite{Marshall2007,Kennedy2008} , applied to the remaining $s-k-1$ points of $\tilde\Gamma$, gives the set of pairs $(t_i,\xi_{t_i})_{i=0}^{s-k-1}$, with $\xi_{t_0}\equiv 0$. The quantity $\xi_{t_i}$ is the driving function at the ``time'' $t_i$ extracted by the algorithm.
\end{enumerate}
A pictorial representation of the described steps is given in Fig.~\ref{fig:zip}
In the hypothesis that the curve results from a $\mathrm{SLE}_\kappa$, the obtained driving function has to be a Brownian process with $\langle\xi^2_t\rangle=\kappa t$. We perform the analysis on the Q and on the T model. The results are given in Fig.~\ref{fig:forzante}. The driving function is found indeed to be a Gaussian process with $\langle\xi_t\rangle=0$ and $\langle \xi^2_t\rangle\sim \kappa t$ with $\kappa\approx 2.1(1)$ (see the Appendix for further details). Although we have not been able to resolve the difference in $\kappa$ between the different lattices, the results strongly support that the obtained curves are $\mathrm{SLE}_\kappa$ with $\kappa\simeq 2.1$.

As a second conformality test, we compute the variance of the winding angle of the curves. We start by picking a random starting edge on the cycle $\mathcal S_{\hat e}$. Then, we choose an orientation at random for the cycle (clockwise or anticlockwise), and we compute a function $\vartheta(e)$ of the cycle edges, in such a way that, if $e+1$ is the subsequent edge along the cycle, $\vartheta(e+1)=\vartheta(e)+\text{angle}(e,e+1)$. Here $\text{angle}(e,e+1)$ is the turning angle from $e$ to $e+1$ measured in radians \cite{wieland2003winding}. We restrict the computation to cycles winding around the system, so that $\media{\vartheta(e)}=0$. The variance of the angle is found to grow with $L$, according to the law 
\begin{equation}
\label{var_winding}
\Var[\vartheta]=a+\frac{\kappa}{4}\ln L,
\end{equation}
see Fig.~\ref{isto}. The values of $\kappa$ are given in Table~\ref{tab:esponenti}. Once again, the T model and the REMP are found to have the same value of $\kappa$. This value is different from the value of $\kappa$ obtained for the Q and the H model and for the REAP. In the hypothesis that the obtained curves are $\mathrm{SLE}_\kappa$, $\Var[\vartheta]$ has to behave as in Eq.~\eqref{var_winding}, and the quantity $\kappa$ appearing in Eq.~\eqref{var_winding} is exactly the parameter of the Schramm--L\"owner evolution \cite{duplantier1988winding, duplantier2002harmonic, wieland2003winding}. The quantity $a$ is instead a nonuniversal constant. Moreover, given an $\mathrm{SLE}_\kappa$, then $D_{\rm f}=\min\left(1+\frac{\kappa}{8},2\right)$ \cite{Rohde2005}. This relation approximately holds for all the considered models, see Table~\ref{tab:esponenti}.

\section{{Conclusions}.} The obtained values $\zeta>0$ are evidence of criticality on the ground state in the RDM on all three different lattices and in the considered random Euclidean matching problems. The scaling exponents of the excitations are found to depend on the bipartiteness of the underlying lattice/graph only (the REMP can be thought as a matching problem on the complete graph, the REAP as a matching problem on a bipartite complete graph). This dependence has been observed also assuming a uniform measure over all possible dimer coverings. Under such assumption, in Refs.~\cite{Moessner2001,Fendley2002}, for example, the T model is found to have short range correlations unlike the same the H and the Q model. Moreover, the fractal dimension of the cycles obtained in the REMP and in the T model is compatible with the one of $2D$ spin-glass domain walls, which is estimated to be $D_{\mathrm f}=1.274(2)$ \cite{Melchert2007, Wang2017, *Wang2018, Corberi_2019,Khoshbakht2018}. These results show that nonbipartite matching problems and spin glasses in $2D$ belong to the same universality class. On the other hand, the H model, the Q model and the REAP have cycles with $D_{\mathrm f}=1.252(2)$, compatible with the fractal dimension of the loop-erased self-avoiding walk in $2D$, which has $D_{\mathrm f}=\frac{5}{4}$ and is a $\mathrm{SLE}_2$ \cite{majumdar1992exact,Lawler2004}. Compatible fractal dimensions have been obtained in Ref.~\cite{Kimchi2018}, where the RDM has been used as a model of $2D$ valence-bond solids. Our result also recover the one obtained in Ref.~\cite{Middleton2000} for defect-induced excitations in the $2D$ elastic medium on the honeycomb lattice. Finally, as test of possible conformal invariance, we mapped the contractible excitations in the half upper plane and we computed the driving function $\xi_t$ generating the process. We found, for all analyzed cases, that $\xi_t$ is a Brownian process with variance $\langle\xi_t\rangle\sim\kappa t$, as expected for an $\mathrm{SLE}_\kappa$. As additional test, the variance of the winding angle is found to scale asymptotically as $\Var[\vartheta]\sim \frac{1}{4}\kappa\ln L$ with $\kappa=8(D_{\rm f}-1)$, as predicted for $\mathrm{SLE}_\kappa$.

Our results uncover a non trivial connection between the RDM on monopartite lattices, the REMP and the EA spin-glass model in $2D$. These models are all critical at zero temperature and, more importantly, share the same universality class, exhibiting signs of conformal invariance. RDMs on bipartite lattices and the REAP, on the other hand, show a similar critical zero-temperature behavior, but with different scaling exponents. The underlying graph topology is therefore an essential feature for the determination of the universality class of the critical point.

\subsection*{Acknowledgments.}The authors are grateful to Carlo Lucibello for providing them his preliminary data about the REMP. The authors would like to thank Bertrand Duplantier, Scott Kirkpatrick, Nicolas Macris and Andrea Sportiello for useful discussions. The research of G.P.~and G.S.~has been supported by the Simons Foundation (Grant No. 454949). R.M. has been supported by Swiss National Foundation Grant No.~200021E 17554. 

\appendix*

\section{Numerical analysis of the driving function.} 
\label{app:num}
We fix $k=60,90,120$ and consider $L=200,400$. We condition the ensemble to $s>s_\mathrm{min}=k+10$ to ensure that $k<s$; however, we checked that the results do not depend appreciably on the particular choice of $s_\mathrm{min}$. 
For each pair $L,k$, we measure the $s-k$ pairs $(t_i,\xi_{t_i})_{i=0}^{s-k-1}$ as described in Section~\ref{sec:conf}, for $\sim 10^5$ independent realizations.

We first performed our analysis on the Q model. The leftmost panels in Fig.~\ref{fig:forzante} show the  ensemble-averaged mean square displacement of $\xi_t$ as a function of time. For small times, up to $t_\mathrm{max}\approx 0.012$, $\left<\xi_t^2\right>$ is approximately linear in $t$.
While the time $t_\mathrm{max}$ does not seem to change appreciably with lattice size, the average number of steps needed to reach it, $\left<s(t_\mathrm{max})\right>$, increases with $L$ and with $k$. Note that the quantity $\left<s(t_\mathrm{max})\right>$ is computed
without taking into consideration the first $k$ steps of the original curve,
which get mapped to the boundary of the domain. For $L=200$ we find
$\left<s(t_\mathrm{max})\right>\approx 110 / 157 / 188$ for $k=60 / 90 / 120$ respectively;
for $L=400$ we find
$\left<s(t_\mathrm{max})\right>\approx 174 / 268 / 325$.
The ratios between $\left<s(t_\mathrm{max})\right>$ and the average number of steps
of the curves are
$0.43,0.49,0.51$ for $L=200$ (again for $k=60,90,120$)
and $0.46,0.58,0.61$ for $L=400$.

To compute $\kappa$ we performed linear fits of $\left<\xi_t^2\right>$ versus $t$,
using only the data at the larger size, $L=400$.
We performed fits in intervals $[0,t_\mathrm{max}]$ with varying $t_\mathrm{max}$.
Fit results are shown in Fig.~\ref{fig:forzante}. The fitted values for the square lattice stabilize around $t_\mathrm{max}\approx 0.01$. The three values of $k$ that we considered yield compatible estimates at the plateau, namely $\kappa=2.07(8)$. %

Besides the scaling of the mean squared displacement, SLE predicts that the normalized process $\xi_t/\sqrt{t}$ is distributed normally with mean $0$ and variance $\kappa$. To have sufficient statistics, we considered the data for $k=60,90,120$ combined, and all times from $t_\mathrm{min}=0.001$ and $t_\mathrm{max}=0.012$ (for smaller values of $t_\mathrm{min}$ lattice artefacts become apparent, affecting the tails of the distribution). Fig.~\ref{fig:forzante} shows that the normalized process is approximately Gaussian as expected. 

Repeating the same procedure for the T model we obtain $\kappa=2.16(13)$. However, notice that, in the T model, the three values of $k$ give slightly inconsistent estimates, see Fig.~\ref{fig:forzante}. The values for $k=60$ do not reach a well defined plateau; excluding them from the overall estimate gives $\kappa = 2.20(8)$.

\bibliography{bibliografia.bib}
\end{document}